\documentclass[prl,showpacs,amsmath,a4paper,twocolumn,superscriptaddress,floatfix]{revtex4-1}
\usepackage{amssymb}
\usepackage{amsmath}
\usepackage{graphicx}
\usepackage{amsfonts}
\usepackage{hyperref}
\usepackage{oldgerm}
\usepackage{color}
\usepackage{bbm}

\newcommand{\bea}{\begin{eqnarray}}
\newcommand{\eea}{\end{eqnarray}}

\newcommand{\ba}{\begin{eqnarray}}
\newcommand{\ea}{\end{eqnarray}}

\DeclareMathOperator{\Tr}{Tr}

\begin{document} 

\title{Witnessing genuine multipartite entanglement with positive maps}

\author{Marcus Huber}
\affiliation{ICFO-Institut de Ciencies Fotoniques, 08860 Castelldefels, Barcelona, Spain}
\affiliation{Universitat Autonoma de Barcelona, 08193 Bellaterra, Barcelona, Spain}
\author{Ritabrata Sengupta}
\affiliation{Department of Mathematical Sciences, Indian Institute of Science Education and Research Mohali, Mohali 140306, India}

\begin{abstract}
We derive a general framework that connects every positive
map with a corresponding witness for partial separability in
multipartite quantum systems. We show that many previous
approaches were intimately connected to the witnesses
derived from partial transposition and that such criteria
can easily be outperformed in higher dimensions by
non-decomposable maps. As an exemplary case we present a
witness that is capable of detecting genuine multipartite
entanglement in bound entangled states.
\end{abstract}
\maketitle
Entanglement is a striking feature of quantum physics that
lies at the very heart of many of its numerous applications
\cite{NC}. Characterizing entanglement is a challenging
task, whose complexity scales very unfavorably with the size
of the system \cite{RevModPhys.81.865, Jens,guth}. In bipartite
systems a major breakthrough in entanglement detection came
with the advent of simple operational criteria for detecting
entanglement in mixed states \cite{H1}. One of the first of
its kind was the famous Peres-Horodecki criterion, also
known as PPT (positivity under partial transposition)
criterion. It provides a method to tell with certainty,
whether two qubit systems are entangled and also provides a
criterion showing whether states can not be distilled from
multiple copies of such states into purer entanglement via
local operations and classical communication (LOCC)
\cite{PhysRevA.53.2046}. Soon after it was realized that one
can exploit the theory of positive, yet not completely
positive maps to obtain complementary and in many cases much
stronger entanglement detection criteria
\cite{PhysRevLett.76.722}. In fact if a state that remains
positive after application of all possible positive maps to
one of its subsystems it is separable with respect to that partition \cite{H1}. 

\par When it comes to multipartite systems the situation becomes
a little more involved. The possible structure behind the
infinitely many decompositions of multipartite quantum
states constitutes an even harder challenge for the
detection of entanglement. Famous instances exemplifying the
complexity of the task are states that are entangled across
every bipartite cut, yet no multipartie entanglement is
necessary to describe them \cite{RevModPhys.81.865} and on the other end of
the spectrum states there are separable with respect to every
bipartite cut, yet they are not completely separable
\cite{B2}. Unfortunately the
paradigmatic tool for entanglement detection in bipartite
systems, positive maps, is an inherently bipartite concept
and applied to multipartite systems it can never reveal more
than mere entanglement across bipartite cuts.\\
Thus entanglement witnesses are the most commonly used tool
 to detect genuine multipartite entanglement in
noisy multipartite quantum systems \cite{guth} and many attempts have been made to frame 
multipartite entanglement detection in a general framework \cite{PhysRevLett.104.210501,WKBKH,mamaju,sv1}. In the bipartite case
there is an intriguing connection between positive maps and
entanglement witnesses, as the latter can be derived from
the former. In this letter we introduce a general framework
that allows to construct witnesses for genuine multipartite
entanglement directly from positive maps. In fact we even
show how any non-partial decomposability can be revealed in
such a way and provide examples where our framework
outperforms the best known witness constructions.\\
To get started let us precisely define the underlying
concepts of separability, positive maps and entanglement
witnesses before we move on to our main theorem.\\ 
A state is considered to be partially separable with respect
to bi-partitions $b\in\mathcal{B}$ if and only if it can be
written as
\begin{align}\label{defmulti}
\rho_\mathcal{B}=\sum_{b\in\mathcal{B}}p_b\left(\sum_i
q_b^i(|\phi_i\rangle\langle\phi_i|)_b\otimes(|\phi_i'\rangle\langle\phi_i'|)_{\overline{b}}\right)\,.
\end{align}
This definition carries the operational meaning of which
resources in terms of separability are required to create
this state via LOCC. A special case are states that are
bi-separable, i.e. $\mathcal{B}$ is the set of all possible
bi-partitions $|\mathcal{B}|= 2^{n-1}-1$. The complement of
the set of bi-separable states is usually referred to as
\emph{genuinely multipartite entangled} states as their
creation via LOCC requires pure states that are not
separable with respect to any partition. Due to the involved
structure of the definition of bi-separability
(\ref{defmulti}) detecting genuine multipartite entanglement
is a challenging task.\\
This is where entanglement witnesses prove useful. These are
self-adjoint operators that have a positive expectation
value for all states $\rho_\mathcal{B}$, while there is at
least one state in the complement for which the expectation
value is smaller than zero. The advantage of witnesses,
while their detection capability is limited to a small
volume of states, is of course the generically easy
experimental access (especially in systems so large that a tomography is nearly impossible as e.g. in Refs. \cite{exp1,exp2}). If global measurements are available a
single measurement is sufficient to reveal entanglement in a
physical system and even for more realistic local
measurements generic witness only require a small fraction
of possible measurements \cite{guth}. Especially for revealing
multipartite entanglement this is a very desirable property
as a full state tomography scales very unfavorably in the
number of systems involved.\\
Positive maps $\Lambda$ (that are not completely positive)
on the other hand constitute a tool for entanglement
detection that require access to the full density matrix and
also the computation of eigenvalues of matrices that are
exponentially large in the number of systems. There is
however a straightforward connection that allows to
construct entanglement witnesses directly from positive
maps, which we will elucidate after some preliminary definitions.\\
 For bipartite systems it is obvious that
\begin{align}
\Lambda_b\otimes\mathbbm{1}_{\overline{b}}[\rho_b]=\sum_i q_i \Lambda[(|\phi_i\rangle\langle\phi_i|)_b]\otimes(|\phi_i'\rangle\langle\phi_i'|)_{\overline{b}})\geq 0\,,
\end{align}
such that any negative eigenvalue after application of the
positive map to the subsystem immediately reveals
entanglement across this bipartition into the subsystem and
its complement. While this can never reveal partial
separability properties in the general sense of
(\ref{defmulti}), positive maps, such as \cite{bru2,hall1,choilam,cho1,kye,os,kossak4} have proven to provide
strong tools in the bipartite case \cite{H1}. There is a
straightforward framework for constructing bipartite
entanglement witnesses from positive maps: If the aim is to
detect a given entangled target state $\sigma$ and there
exists a positive map $\Lambda$, such that
$\Lambda\otimes\mathbbm{1}[\sigma]$ has at least one
negative eigenvalue with corresponding eigenvector
$|n\rangle$, then
\begin{align}
W_\Lambda=\Lambda^*\otimes\mathbbm{1}[|n\rangle\langle n|]\,,
\end{align}
where $\Lambda^*$ is the dual of the positive map $\Lambda$,
constitutes an entanglement witness that will detect the
state $\sigma$ to be entangled. Such a procedure is of
course very helpful in experimental entanglement
verification if one has a reasonable guess what the state of
the system under investigation should be. Then one can apply
this procedure and end up with an experimentally feasible
witness operator that should be able to reveal entanglement
in the system. Unfortunately this procedure only works for
the bipartite case as the application of a map on a system
necessarily implies a bi-partition. Now we continue with the
main result of our paper, where we present a framework that
enables such a construction also for partial separability
and thus for genuine multipartite entanglement. We start
directly with the main theorem:\\

\textbf{Theorem:} \textit{For any positive map $\Lambda$ and
set of partitions of a multipartite state $\mathcal{B}$, the
following expression is always positive for mixed states
$\rho$, which can be decomposed into pure states that are
separable with respect to any of the partitions in
$\mathcal{B}$}
\begin{align}
\text{Tr}\left[\rho\left(\sum_{b\in\mathcal{B}}\tau_b+Q\right)\right]\geq 0
\end{align}
where we have used the abbreviated notation $Q=N+P$,
$P=\sum_{\eta,\eta'}|\eta\rangle\langle\eta'|\max[0,\min_{b\in\mathcal{B}}[\Re e[\langle\eta|\Lambda^*_b\otimes\mathbbm{1}_{\overline{b}}[|\psi_b\rangle\langle\psi_b|]|\eta'\rangle]]]$,
$N=\sum_{\eta,\eta'}|\eta\rangle\langle\eta'|\min[0,\max_{b\in\mathcal{B}}[\Re e[\langle\eta|\Lambda^*_b\otimes\mathbbm{1}_{\overline{b}}[|\psi_b\rangle\langle\psi_b|]|\eta'\rangle]]]$
and
$\tau_b=[\Lambda^*_b\otimes\mathbbm{1}_{\overline{b}}[|\psi_b\rangle\langle\psi_b|]-Q]_+$
(with $[A]_+$ we denote the non-negative part of the
spectrum of $A$, i.e. we project onto the  eigenspace
spanned by eigenvectors belonging to positive
eigenvalues).\\
\textbf{Proof}:\\
The first observation required is the fact that
\begin{align}
\text{Tr}[\rho(\Lambda^*_b\otimes\mathbbm{1}_{\overline{b}}[|\psi_b\rangle\langle\psi_b|]\geq 0\,\forall\,\rho\in\mathcal{S}_{b|\overline{b}}
\end{align}
which is really just a restatement of the fact that $\Lambda$ is a positive map. Next we point out that
\begin{align}
\text{Tr}[\rho [A]_+]\geq\text{Tr}[\rho A]\,,
\end{align}
and thus
\begin{align}
\text{Tr}[\rho(\tau_b+Q)]\geq \text{Tr}[\rho(\Lambda^*_b\otimes\mathbbm{1}_{\overline{b}}[|\psi_b\rangle\langle\psi_b|]
\end{align}
Now if we write down a state that is decomposable into states $\rho_b$, separable with respect to bipartition in $\mathcal{B}$ as
\begin{align}
\rho_{\mathcal{B}}=\sum_b p_b \rho_b
\end{align}
we find that
\begin{eqnarray}
&&\Tr\left[\rho\left(\sum_{b\in\mathcal{B}}\tau_b+Q\right)\right]=\nonumber \\
&&\sum_{b\in\mathcal{B}}p_b\text{Tr}[\rho_b(\tau_b+Q)]+\sum_{\{b'\neq b\}\in\mathcal{B}}p_b'\text{Tr}[\rho_{b'}(\tau_b)]\geq0
\end{eqnarray}
which completes the proof.\\
The idea behind the theorem is straightforward and actually is far more general than just referring to positive maps: If one has a set of witnesses for detecting entanglement across different partitions that have some overlapping matrix elements collected in $Q$, one can separate every witness $W_b=Q+M_b$ and by making sure that $M_b$ is positive semi-definite (as we did in our theorem by using only the positive part of the spectrum) it immediately follows that $W_{GME}=Q+\sum_b M_b$ is a witness for genuine multipartite entanglement. The specific role of positive maps in our theorem is actually just to open the possibility for using non-decomposable witnesses in this context and to maximize the ovelap (i.e. the norm of $Q$) in a natural way. The difficulty for using generic maps here is of course
finding suitable "`witness states"' $|\psi_b\rangle$ that
maximize the overlap of negative elements for different
partitions, i.e. the operator norm of $Q$. Using partial
transpose as an exemplary criterion here we can illustrate
the method in the three qubit case:\\
\textit{Example 1}\\
If we want to detect genuine multipartite entanglement in a three qubit GHZ state $|GHZ\rangle=\frac{1}{\sqrt{2}}(|000\rangle+|111\rangle))$, through a witness derived from the PPT criterion, the choice of $|\psi_b\rangle$ is quite straightforward. If we choose
\begin{align}
|\psi_1\rangle=\frac{1}{\sqrt{2}}(|011\rangle-|100\rangle)\nonumber\\
|\psi_2\rangle=\frac{1}{\sqrt{2}}(|101\rangle-|010\rangle)\nonumber\\
|\psi_3\rangle=\frac{1}{\sqrt{2}}(|110\rangle-|001\rangle)
\end{align}
we end up with a witness operator
\begin{align}
W=\frac{1}{2}\mathbbm{1}-|GHZ\rangle\langle GHZ|\,,
\end{align}
which is well known \cite{guth} and even necessary and
sufficient for detecting multipartite entanglement in
GHZ-diagonal states \cite{guse}. Indeed
Refs.\cite{PhysRevLett.104.210501,WKBKH,mamaju} introduce a
framework for multipartite entanglement detection, whose
linearized version is exactly corresponds to our main
theorem here. Using this framework one can always find the
corresponding $|\psi_b\rangle$ for detecting multipartite
entanglement. however as we show this framework (and the
multipartite entanglement witnesses derived from the PPT
criterion) can easily be outperformed by a simpler choice of
maps. For instance the Breuer-Hall map \cite{bru2,hall1},
Choi's map \cite{choilam} and its  generalizations 
\cite{kye, os}, have the advantage that the negative
off-diagonal elements in
$\Lambda^*_b\otimes\mathbbm{1}_{\overline{b}}[|\psi_b\rangle\langle\psi_b|]$
generically correspond to the off diagonal elements of
$|\psi_b\rangle$. This facilitates the search immensely as
for the theorem to maximize detection strength we require
the off diagonal elements of
$\Lambda^*_b\otimes\mathbbm{1}_{\overline{b}}[|\psi_b\rangle\langle\psi_b|]$
to be as similar as possible. For such maps where this is
naturally the case it is sufficient for detecting a target
state $|\Psi_t\rangle$ to choose
$|\psi_b\rangle=|\Psi_t\rangle\,\forall\,b$ (or if one is
lucky one and finds an eigenvector of
$|\psi_b\rangle=|\Psi_t\rangle\,\forall\,b$ in all
partitions with a negative sign and high modulus, then this
is of course the obvious choice). We will now present an
example (similar in spirit to one example in
Ref.~\cite{acin}) where this advantage becomes immediately
evident:\\
\textit{Example 2}\\
Adopting the following short hand notation for GHZ like states
\begin{align}
|\alpha,x,y,\lambda_b\rangle=\prod_{i\in\alpha}\sigma_i\otimes\mathbbm{1}_{\overline{i}}(\sqrt{\lambda_\alpha}|x\rangle^{\otimes n}+\sqrt{\lambda_\alpha^{-1}}|y\rangle^{\otimes n})\,,
\end{align}
with $\sigma=|x\rangle\langle y|+|y\rangle\langle x|$. And the corresponding projector we denote as $P_\alpha(x,y,\lambda)$, such that we can introduce the operator
\begin{align}
E(\{\lambda_\alpha\})=3|GHZ_3\rangle\langle GHZ_3|+\sum_{i=1,2,3}\sum_{x<y}\sum_{y=1,2} P_i(x,y,\lambda)\,,
\end{align}
where $|GHZ_3\rangle=\frac{1}{\sqrt{3}}(|000\rangle+|111\rangle+|222\rangle)$ and with this finally the density operator
\begin{align}
\rho(\{\lambda_\alpha\})=\frac{E(\{\lambda_\alpha\})}{\text{Tr}(E(\{\lambda_\alpha\}))}\,.
\end{align}
It is immediately evident that this density matrix is
invariant under partial transposition (since the off diagonal elements of $|GHZ_3\rangle$ after partial transposition  in system $b$ are the same as $|\alpha,x,y,\lambda_b\rangle$), so the PPT criterion is not even able to reveal bipartite entanglement in this system. However using Choi's map for $d=3$:
\begin{equation}
((a_{ij}))\mapsto
\frac{1}{2} 
\begin{pmatrix}
a_{11}+a_{33} & -a_{12} & -a_{13}\\
-a_{21} & a_{22}+a_{11} & -a_{23} \\
-a_{31} & -a_{32} & a_{33}+a_{22}
\end{pmatrix}\,,
\end{equation} one can easily check that
this state is indeed PPT entangled (i.e. definitely bound
entangled) across every bipartition for values of
$0<\lambda_\alpha<1$. An immediate implication is the fact
that if the state is multipartite entangled it can not be
detected by our theorem using the PPT and also not from the
techniques developed in
Refs.~\cite{guse,PhysRevLett.104.210501,WKBKH,mamaju}. Using
the very simple and straightforward choice
$|\psi_1\rangle=|\psi_2\rangle=|\psi_3\rangle=|GHZ_3\rangle$,
$\lambda_\alpha=\lambda\,\forall\,\alpha$ and again Choi's
map we can directly apply our method to check partial
separability properties. We find that the witness is
violated for all values of $\lambda$ between $0$ and
$\frac{1}{3}$ and thus this bound entangled state is indeed
genuinely multipartite entangled. Other examples were found for symmetric states in Refs.~\cite{GMEbound1,GMEbound2,GMEbound3,GMEbound4}) and in Ref.~\cite{Marco} the authors construct a different framework that allows for the construction of PPT-GME states, however to our knowledge that is the first explicit example in a $3\otimes3\otimes3$ system.\\
 The violation of this witness is even so significant that it exhibits a notable
noise robustness with respect to white noise. 

\begin{figure}[tp!]
\begin{center}
  \includegraphics[scale=1]{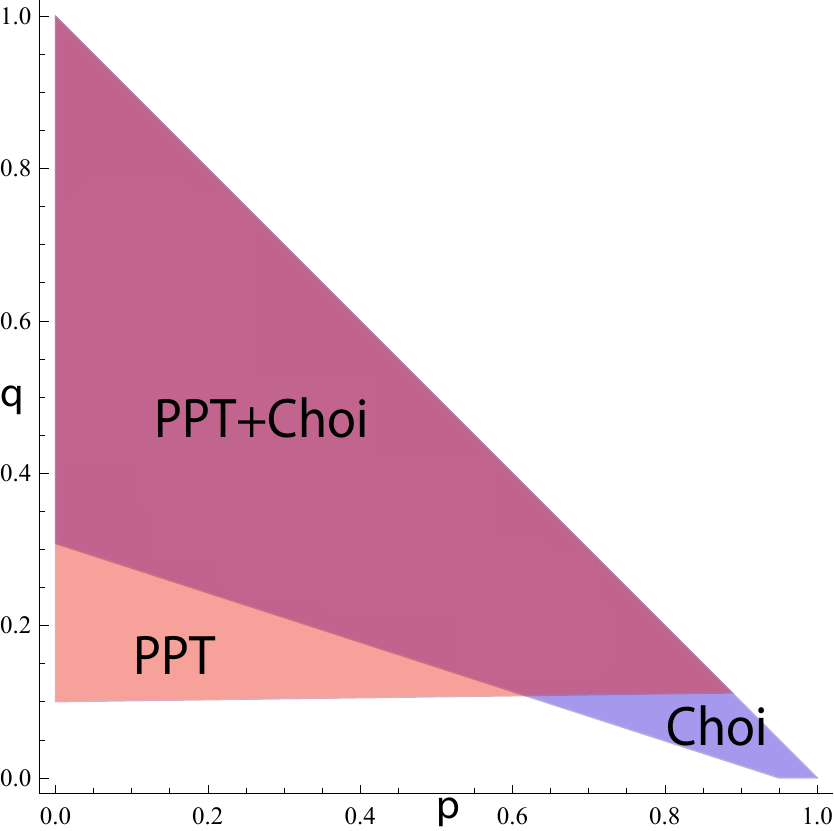}
  \caption{\label{Example3} Here we illustrate the detection
power for state (\ref{state3}) of two different maps. The
regions are labeled according to which criterion detects the
state for these values of $p$ and $q$ to be genuinely
multipartite entangled (PPT refers to partial transposition
and Choi to Choi's map).}
\end{center}
\end{figure}

If we mix the state, e.g. for a choice of $\lambda=\frac{1}{9}$, with the maximally mixed state, i.e.
\begin{align}
\rho_{noise}(p)=p\frac{\mathbbm{1}}{27}+(1-p)\rho
\left(\frac{1}{9} \right)\,,
\end{align}
we find that the white noise resistance, i.e. the critical
value of white noise admixture $p$, until which genuine
multipartite entanglement can still be detected is
$p_{crit}=\frac{9}{179}\approx 5\%$.\\
We have just shown that Choi's map provided an
advantage for specific states, while in general different maps
cannot be considered superior in terms of their
entanglement detection strength (even with the optimal
choice of $Q$ for a given class of states). To illustrate
this point let us consider the following two-parameter
family of states
\begin{align}\label{state3}
\rho_{example}=p |GHZ_3\rangle\langle GHZ_3|+q
\rho\left(\frac{1}{9}\right) +\frac{1-p-q}{27}\mathbbm{1}\,,
\end{align}
and use our theorem to construct the witnesses in the same
fashion as in the first two examples. The results are
illustrated in Figure~(\ref{Example3}) and showcase the
detection power of the witnesses derived from our main
theorem and straightforward choices of $|\psi_b\rangle$
without any optimization involved.\\

In conclusion we presented a framework that directly connects
positive maps with witnesses for partial separability. The
construction is simple, operational and experimentally
friendly. We illustrated the power of the criterion by
presenting the first example of a $3\otimes 3\otimes 3 $
bound entangled that is at the same time genuinely
multipartite entangled. We expect that this method should
find applications in all tasks that aim at characterizing
multipartite entanglement. In the future it will be of
interest to study the connection to semidefinite
characterizations of supersets of partially separable states
(as with the "`PPT mixers"' from Refs.~\cite{gu3,gu4}). One
intriguing question that is left open is the general
strength of such criteria---while in the bipartite case it
is obvious that such a witness construction is capable of
detecting all entangled states, the case is not so clear in
the multipartite case. Can we construct multipartite
entangled states that in principle cannot be detected by our
framework?\\
\textit{Acknowledgments} We would like to thank Otfried
G\"uhne, Claude Kl\"ockl, Tobias Moroder and Sabine W\"olk for inspiring
discussions and especially Marco Piani for important comments. M.H. acknowledges funding from the MarieCurie
grant N302021 "Quacocos". R.S. wants to thank Andreas
Winter,  Marek K\'us for useful discussions  and the Center for
Theoretical Physics of the Polish Academy of Sciences, where
 a considerable portion of this work was materialised.


\begin{thebibliography}{99}

\bibitem{NC}%
  M.~A. Nielsen and I.~L. Chuang, \emph{Quantum computation and quantum information}, \href {\doibase 10.2277/0521635039}{Cambridge University Press}, {Cambridge}, (2000).
\bibitem{RevModPhys.81.865}  R. Horodecki, P. Horodecki, M. Horodecki, and K. Horodecki, \href{\doibase 10.1103/RevModPhys.81.865 }{Rev. Mod. Phys. {\bf 81}, 865} (2009).
\bibitem{guth} O. G\"uhne, G. Toth, \href{\doibase 10.1016/j.physrep.2009.02.004}{Physics Reports 474, 1} (2009). 
\bibitem{Jens} C. Eltschka, J. Siewert, \href{http://arxiv.org/abs/1402.6710}{arXiv:1402.6710} (2014).
\bibitem{H1} M. Horodecki P. Horodecki, R. Horodecki, \href {\doibase 10.1016/S0375-9601(96)00706-2}{Phys. Lett. A {\bf 223}, 1} (1996).
\bibitem {PhysRevA.53.2046} C.~H. Bennett, H.~J. Bernstein, S. Popescu, B. Schumacher, \href{\doibase 10.1103/PhysRevA.53.2046}{Phys. Rev. A {\bf 53}, 2046} (1996).
\bibitem {PhysRevLett.76.722} C. H. Bennett, G. Brassard, S. Popescu, B. Schumacher, J. A. Smolin, W. K. Wootters, \href{\doibase 10.1103/PhysRevLett.76.722}{Phys.
  Rev. Lett. {\bf 76}, 722} (1996).
\bibitem{B2}C.H. Bennett, D.P. DiVincenzo, T. Mor, P.W. Shor, J.A. Smolin, B.M. Terhal, \href {\doibase  10.1103/PhysRevLett.82.5385}{Phys.Rev.Lett. {\bf 82}, 5385} (1999).
\bibitem{PhysRevLett.104.210501} M. Huber , F. Mintert, A. Gabriel and B.C. Hiesmayr, \href{\doibase  	10.1103/PhysRevLett.104.210501}{Phys.Rev.Lett. {\bf 104}, 210501} (2010).
\bibitem{WKBKH} J.-Y.~Wu, H.~Kampermann, D.~Bru\ss, C.~Kl\"ockl, and M.~Huber, \href{\doibase  	10.1103/PhysRevA.86.022319}{Phys. Rev. A {\bf 86}, 022319} (2012).
\bibitem{mamaju} M. Huber, M. Perarnau-Llobet, J. I. de Vicente, \href{\doibase  	10.1103/PhysRevA.88.042328}{Phys. Rev. A {\bf 88}, 042328} (2013).
\bibitem{sv1} J. Sperling, W. Vogel, \href{\doibase 10.1103/PhysRevLett.111.110503}{Phys. Rev. Lett. 111, 110503} (2013).
\bibitem{exp1} T. Monz \textit{et. al.}, \href{\doibase	10.1103/PhysRevLett.106.130506}{Phys. Rev. Lett. 106, 130506} (2011).
\bibitem{exp2} M. Krenn \textit{et. al.}, \href{\doibase 10.1073/pnas.1402365111}{Proc. Natl. Acad. Sci. {\bf 111} (17) 6122-6123} (2014).
\bibitem{bru2} H.-P. Breuer, \href {\doibase 10.1103/PhysRevLett.97.080501}{Phys. Rev. Lett. {\bf 97}, 080501} (2006).
\bibitem{hall1} W. Hall, \href{\doibase 10.1088/0305-4470/39/45/020}{J. Phys. A {\bf 39}, 14119} (2006).
\bibitem{choilam} M.~D. Choi and T.~Y. Lam, \href {\doibase 10.1007/BF01360024}{Math. Ann. {\bf 231}, 1} (1977/78)
\bibitem{cho1} S.~J. Cho, S.-H. Kye and S.~G. Lee, \href{\doibase 10.1016/0024-3795(92)90260-H}{Linear Algebra Appl. {\bf 171}, 213} (1992).
\bibitem{kye} S.-H. Kye, {\emph{Elementary operators and applications} }, World Sci. Publ., River Edge, NJ, {205--209} (2009).
\bibitem{os} H. Osaka, Publ. Res. Inst. Math. Sci. {\bf 28}, 747 (1992).
\bibitem{kossak4} D. Chru{\'s}ci{\'n}ski and A. Kossakowski, \href {\doibase  10.1007/s11080-007-9052-4}{Open Syst. Inf. Dyn. {\bf 14}, 275} (2007); D. Chru{\'s}ci{\'n}ski and G. Sarbicki, \href{http://xxx.lanl.gov/abs/1402.2413}{arXiv:1402.2413} (2014).
\bibitem{guse} O. G\"uhne, M. Seevinck, \href{\doibase 10.1088/1367-2630/12/5/053002}{New J. Phys. {\bf 12}, 053002} (2010).
\bibitem{acin} A. Acin, D. Bruss, M. Lewenstein and A. Sanpera, \href{\doibase 10.1103/PhysRevLett.87.040401 }{Phys. Rev. Lett. {\bf 87}, 040401} (2001).
\bibitem{GMEbound1} G. Toth and O. G\"uhne, \href{\doibase  	10.1103/PhysRevLett.102.170503}{Phys. Rev. Lett. {\bf 102}, 170503} (2009).
\bibitem{GMEbound2} G. Toth and O. G\"uhne, \href{\doibase  	10.1007/s00340-009-3839-7}{Appl. Phys. B {\bf 98}, 617} (2010).
\bibitem{GMEbound3} J. Tura, R. Augusiak, P. Hyllus, M. Kus', J. Samsonowicz, M. Lewenstein, \href{\doibase 10.1103/PhysRevA.85.060302 }{Physical Review A {\bf 85}, 060302(R)} (2012).
\bibitem{GMEbound4} R. Augusiak, J. Tura, J. Samsonowicz, M. Lewenstein, \href{\doibase  	10.1103/PhysRevA.86.042316}{Physical Review A {\bf 86}, 042316} (2012).
\bibitem{Marco} M. Piani and C. Mora, \href{\doibase 10.1103/PhysRevA.75.012305}{Phys. Rev. A {\bf 75}, 012305} (2007).
\bibitem{gu3} B. Jungnitsch, T. Moroder, O. G\"uhne, \href{\doibase 10.1103/PhysRevLett.106.190502}{Phys. Rev. Lett. {\bf 106}, 190502} (2011).
\bibitem{gu4} T. Moroder, J.-D. Bancal, Y.-C. Liang, M. Hofmann, O. G\"uhne, \href{\doibase 10.1103/PhysRevLett.111.030501}{Phys. Rev. Lett. {\bf 111}, 030501} (2013).
\end{thebibliography}
\end{document}